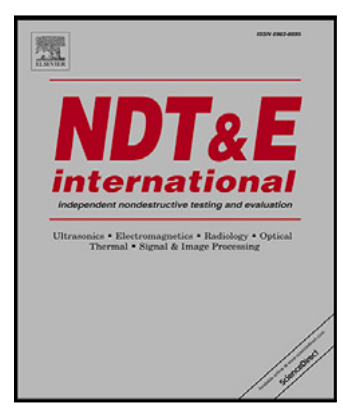

Research Paper

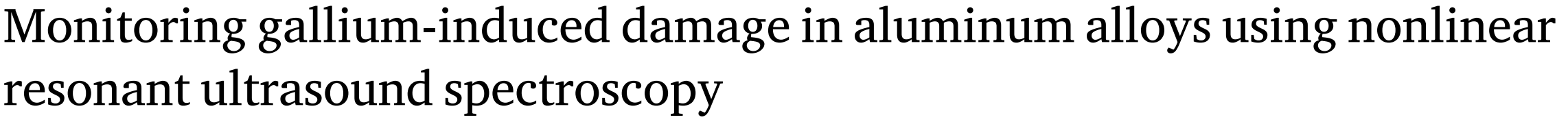

# Monitoring gallium-induced damage in aluminum alloys using nonlinear resonant ultrasound spectroscopy

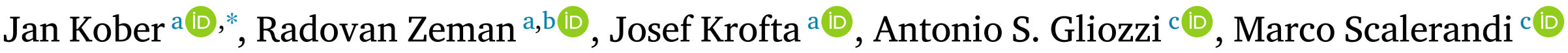

Jan Kober [a,*], Radovan Zeman [a,b], Josef Krofta [a], Antonio S. Gliozzi [c], Marco Scalerandi [c]

[a] Institute of Thermomechanics, Czech Academy of Sciences, Prague, Czechia

[b] Faculty of Nuclear Sciences and Physical Engineering, Czech Technical University in Prague, Czechia

[c] DISAT, Condensed Matter Physics and Complex Systems Institute, Politecnico di Torino, Italy

ABSTRACT

Nonlinear Resonant Ultrasound Spectroscopy is a nonlinear ultrasonic technique which allows monitoring small variations in the microstructure of a medium and thus allows materials characterization and monitoring of damage evolution. Application of the technique to monitor Liquid Metal Embrittlement induced by gallium penetration in aluminum is presented here. To define indicators of material degradation, data treatment using the Singular Value Decomposition approach is introduced and discussed. Experimental results show that nonlinear properties are correlated with the state of the liquid metal in the solid matrix, allowing to identify different phases in the process of gallium diffusion along grain boundaries and within the bulk of individual grains. Furthermore, the evolution of gallium damage allows to study correlations between nonlinear, fast and slow dynamic properties.

## 1. Introduction

Exposure of metal alloys to a liquid metal environment can lead to a specific type of mechanical degradation called Liquid Metal Embrittlement (LME). Although the conditions for the occurrence of LME depend on the particular metal alloy–liquid pairing, the common result is that LME causes a serious decrease in material capacity for plastic deformation [1] and at the same time increases the crack propagation rate, which can lead to rapid structural failure [2]. The risks of LME are present in the aerospace, automotive and power industries [3].

In this paper, we focus on the gallium-induced damage in aluminum alloys. The penetration of gallium into the microstructure is a complex process in which the liquid metal travels along the grain boundaries [4]. The specificity of such damage is that there is not always internal cracking, while the degradation of mechanical properties is mainly due to the presence of gallium in the intergrain volume [5]. The microstructure of gallium damaged alloys resembles, to some extent, a consolidated granular material where crystalline grains are surrounded by a softer intergrain interface.

In order to follow the effect of gallium embrittlement on elastic properties, ultrasonic methods are suitable. Gallium penetration has been shown to affect the linear elastic modulus: a decrease in the velocity of the elastic wave has been observed [6]. Nonlinear ultrasonic methods have never been applied to monitor LME, despite being in general advantageous, due to the higher sensitivity to sub-wavelength defects [7,8] and to their capability to assess the microstructural state of consolidated granular materials in general [9]. Among the various nonlinear methods, Nonlinear Resonant Ultrasound Spectroscopy (NRUS) is particularly promising as it provides information on linear and nonlinear elastic properties and damping of materials at the same time [10–13], which have been shown to provide complementary information [14–19].

The standard NRUS measurement protocol uses sine-train probing at fixed set of frequencies while the excitation amplitude is increased in steps. Due to elastic nonlinearity, a shift of the resonant frequency is observed as a function of the excitation strain, consequence of the velocity change. Experimental results obtained comparing NRUS and Dynamic Acoustoelastic Testing measurements [20] demonstrated that the effects of slow and fast dynamics are an integral part of NRUS probing. The dynamic excitation of NRUS probing is often sufficient to induce conditioning of the material, i.e., the evolution in time of linear modulus and damping towards a non-equilibrium steady state when the sample is subject to a dynamic strain at a given amplitude. Since the recovery of the induced velocity change is a long-time process, the slow dynamic effects accumulate during NRUS probing (i.e., during successive probing at increasing amplitudes) [21].

When the linear elastic response changes in time due to an unstable environment or a damaging process, e.g. when long time monitoring is performed, it is necessary to separate the frequency shift caused by the nonlinearity and the temporal evolution of the linear modulus. For this purpose, a novel method is proposed for processing NRUS data. First, it is possible to significantly reduce the time of NRUS measurements by using swept frequency excitations (chirps) instead of sine trains [13], which contributes to the temporal resolution of the method. Second, using baseline measurements, it is possible to separate not only nonlinear and environmental effects, but also a slow dynamic contribution from the total frequency shift. Using data-based analysis of our time-series data, instead of relying on parametric functions, allows us to simplify the outputs of the NRUS monitoring without sacrificing their complexity. For this purpose, the Singular Value Decomposition (SVD) is used to identify key components (singular vectors) of both the strain and the temporal dependences of NRUS data. As a result, the evolution of gallium-induced damage could be followed using indicators for linear and nonlinear moduli and damping, and slow and fast dynamics. Furthermore, since the gallium damaging process spans several hours, it offers a unique opportunity to study correlations of nonlinear indicators in a wide range of magnitudes.

* Corresponding author.
*E-mail address:* kober@it.cas.cz (J. Kober).

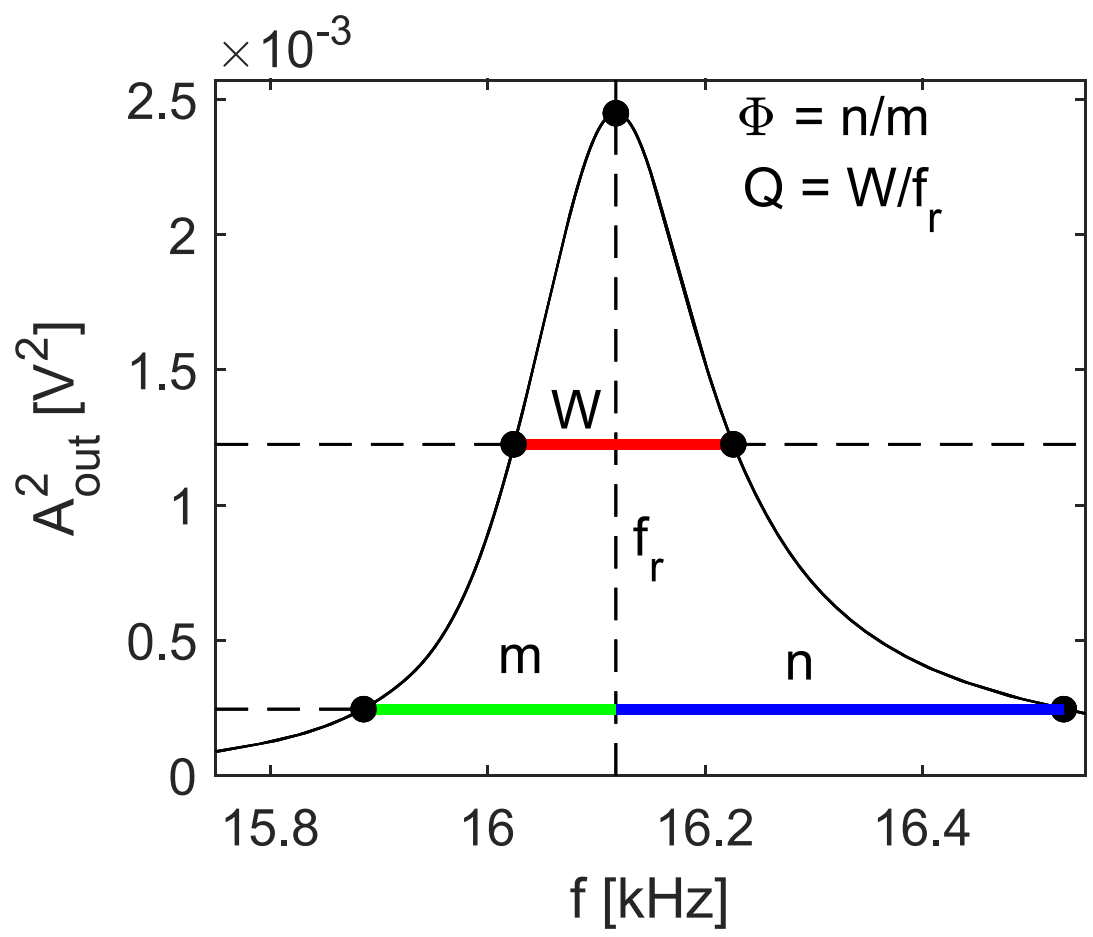

**Fig. 1.** Definition of NRUS parameters. Schematic power spectrum of the resonance peak. The $Q$ factor is determined from the resonance frequency and full width at half maximum $W$ and the asymmetry factor $\Phi$ as a ratio of half widths $m$ to $n$ at 1/10 of the peak height. Experimental results shown correspond to data taken after one hour from the beginning of gallium embrittlement and at low excitation amplitude.

The paper is organized as follows. The theoretical background for the data treatment and separation of NRUS data is given in Section 2, together with the introduction of the specific protocol for NRUS tracking and for data processing. Section 3 contains information on materials and damaging protocol, experimental setup and measurement data processing. The penetration of gallium and its effects on the microstructural and mechanical properties are discussed there as well. The results of gallium damaging monitoring are given in Section 4, for both amplitude and temporal dependencies. Section 5 contains a discussion of the results focusing on parameters correlations. Detailed discussion of data processing (including Singular Value Decomposition approach), repeated experiments and efficiency of baseline correction are included in the Appendices.

## 2. Nonlinear resonant ultrasound spectroscopy

### 2.1. Definition of measured quantities

Resonance experiments have been demonstrated to be a useful tool for assessing the viscoelastic properties of materials. They are based on probing the material with elastic waves at low excitation amplitudes and sweeping frequency in a fixed range. The amplitude of the detected signal defines the resonance curve, see Fig. 1, and the resonant frequency $f_r$ is proportional to the elastic wave velocity $c$, which in turn is a function of the modulus and density of the material. The resonance frequency divided by the width at half height of the resonance curve defines the $Q$ factor, which is related to the damping coefficient.

In the case of nonlinear elastic materials, resonance frequency and damping are dependent on the amplitude of excitation, which is the basis for Nonlinear Resonant Ultrasound Spectroscopy (see Fig. 2). Resonance curves are probed at different drive amplitudes bringing information on the nonlinearity of elastic modulus, which is contained in the shift of the resonance frequency with increasing probing strain:

$$\delta c(\varepsilon) = \frac{c(\varepsilon) - c_0}{c_0} = \frac{f_r(\varepsilon) - f_0}{f_0}, \tag{1}$$

where $c_0$ is the linear velocity (i.e., that measured at the lowest amplitude of the sequence) and $c(\varepsilon)$ is the velocity measured at loading strain $\varepsilon$. In the following, we will continue to refer to elastic wave velocity rather than resonant frequency, as it is a property of the material and not of a particular sample like the latter. Note that the global velocity change observed in NRUS is a result of a distribution of local velocity changes that depend on the excitation mode [20].

As recent results have shown, fast and slow dynamic effects emerging from bond breaking, friction or fluid redistribution can play a crucial role in the nonlinear response of consolidated granular and/or damaged materials [21]. Within the duration of a single NRUS experiment, external/lab conditions (e.g. humidity or temperature) might slightly vary, thus influencing wave velocity as well. We can write the total relative change of elastic wave velocity observed in NRUS as follows:

$$\delta c(\varepsilon, t) = \delta c_{NL}(\varepsilon) + \delta c_{SD}(\varepsilon, t) + \delta c_{EX}(t), \tag{2}$$

where $\delta c_{NL}$ is an instantaneous velocity change due to elastic nonlinearity and fast dynamics, $\delta c_{SD}$ is a persistent velocity change induced by probing evidenced by slow dynamic relaxation, $\delta c_{EX}$ is a possible change of linear wave velocity due to external factors such as temperature variations or increasing damage (which affects the linear modulus and therefore does not depend on the excitation strain $\varepsilon$). The nonlinear component $\delta c_{NL}$ is dependent on the excitation strain and is generated on a very short time scale which is shorter/comparable to the probing wave period. As such, it is fully repeatable under stationary experimental conditions. The slow dynamic component $\delta c_{SD}$ depends on the duration of probing, the probing history and includes cumulative conditioning effects (hence its time dependence). It builds up in a long time, much longer than the wave period and relaxes very slowly. The velocity change due to external factors $\delta c_{EX}$ is only relevant if the scale and the rate of linear modulus change within the duration of the experiment is comparable with the observed nonlinear or slow dynamic effects.

In addition to assessing the resonant frequency, the shape of the resonant peak also carries information on the material response. Damping is related to the width of the resonance peak, and fast dynamics is responsible for the observed peak asymmetry [22,23]. See Fig. 1 for the definitions of related parameters, the quality factor $Q$ and the peak asymmetry $\Phi$. Similar considerations and equations introduced for velocity are valid also for $Q$ and $\Phi$.

### 2.2. Optimizing for tracking and separation

The NRUS measurement procedure adopted here is designed to allow separation of fast, slow and external contributions to the total velocity change and, at the same time, to offer sufficient temporal resolution for tracking the ongoing damaging process.

The temporal resolution is achieved by keeping the measurement time to the minimum. Instead of probing the resonant peak by a number of fixed frequency sine-train excitations [21,24], a chirp excitation allows to sweep over the resonance peak in a fraction of time. Performing the measurements fast enough also helps in the observation and separation of slow dynamic effects, as discussed later.

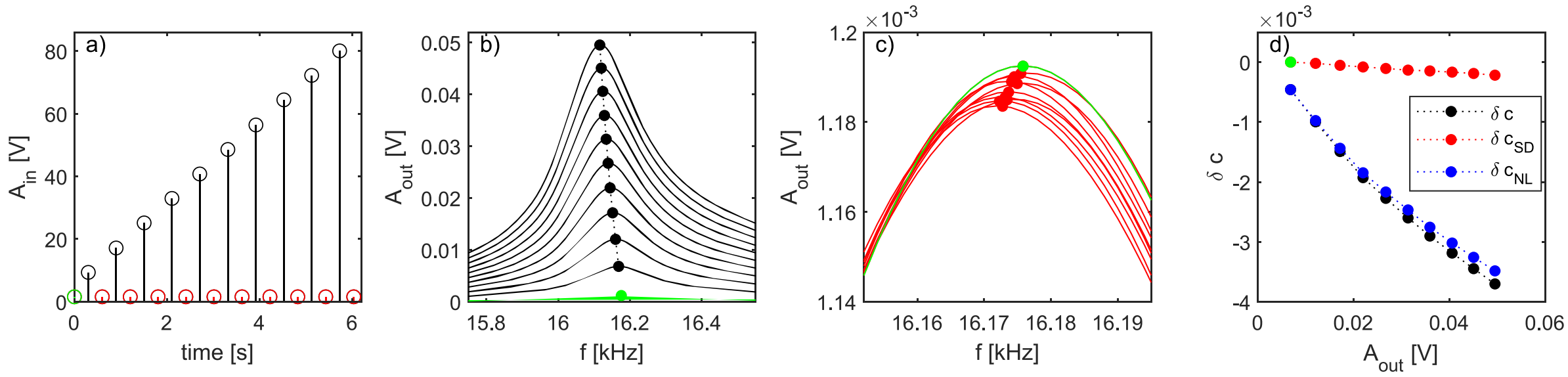

**Fig. 2.** NRUS measurement. (a) Amplitude sequence and timing of the NRUS probing. Each amplitude step (in black) is followed by a low-amplitude baseline measurement (in red). The first low-amplitude measurement (in green) tracks the undisturbed linear state $c_0$. (b) Amplitude spectra of NRUS sequence shows a decrease of resonant frequency with amplitude. (c) A close-up view of the development of baseline measurements showing the decrease of resonant frequency. (d) Amplitude dependence of the nonlinear and slow dynamic components to the relative change of velocity (see Eq. (2) and (3)). Experimental results shown in (b) to (d) correspond to data taken after one hour from the beginning of gallium embrittlement. (For interpretation of the references to color in this figure legend, the reader is referred to the web version of this article.)

The separation of different contributions in Eq. (2) in a particular NRUS experiment is possible by using a specific excitation amplitude sequence, in which a baseline measurement is introduced [21,25,26]. The protocol consists in an alternating excitation amplitude sequence, i.e., after each amplitude step a measurement is done at the lowest excitation amplitude (see Fig. 2a). As mentioned, from the resonance spectra obtained at increasing/baseline excitation amplitudes (see Fig. 2b and c), the resonance frequency can be determined and the velocity derived: the spectra at increasing amplitudes allow us to derive $c(\varepsilon, t)$, while the spectra at the baseline excitation amplitude, successive to the excitation at amplitude $\varepsilon$, give the baseline velocity $c_{\mathrm{BL}}(\varepsilon, t)$ (Fig. 2c). Both decrease (softening) with increasing strain.

The first baseline amplitude probing of the NRUS sequence represents the linear wave velocity $c_0$ in an undisturbed state (assuming a sufficient delay has elapsed from the previous NRUS measurement to ensure the slow dynamics component has relaxed). All subsequent baseline measurements of $c_{\mathrm{BL}}$ contain contributions from slow dynamics and external factors at the same time, the latter of which must be removed. To address this, consider that NRUS probing conducted at successive times yields a sequence of linear velocities $c_0(t)$ over the entire monitoring period. From this sequence, the evolution of the external contribution $c_{\mathrm{EX}}(t)$ can be approximated. In this study, a linear interpolation (in time) was employed. Even though more accurate estimate was possible, it was not needed in the present context.

We can now rewrite Eq. (2) explicitly, keeping the order of contributions, as

$$\begin{aligned} \delta c(\varepsilon, t) = \frac{c(\varepsilon, t) - c_0}{c_0} &= \frac{c(\varepsilon, t) - c_{\mathrm{BL}}(\varepsilon, t)}{c_0} \\ &+ \frac{c_{\mathrm{BL}}(\varepsilon, t) - c_{\mathrm{EX}}(t)}{c_0} + \frac{c_{\mathrm{EX}}(t) - c_0}{c_0}, \end{aligned} \tag{3}$$

where $c_0$ is the linear wave velocity in an undisturbed state and $c_{\mathrm{EX}}$ tracks the velocity evolution due to external factors within the duration of the measurement sequence. Since timing is accurately controlled in the experiment, $c_{\mathrm{EX}}$ can be rewritten as a function of conditioning strain amplitude, $c_{\mathrm{EX}}(t) = c_{\mathrm{EX}}(t(\varepsilon)) = c_{\mathrm{EX}}(\varepsilon)$, where $t(\varepsilon)$ denotes the time at which the strain amplitude $\varepsilon$ is probed.

Note that a similar approach could be introduced to separate slow and fast dynamic effects on the $Q$ factor and peak asymmetry $\Phi$, but it has not been considered in this study due to the presence of excessive noise in the respective baselines, which rendered them unsuitable for analysis.

### 2.3. Describing the nonlinear amplitude dependence

Part of the challenge in coherently describing the evolution of nonlinearity during a damaging process is in the description of the amplitude dependence of the evaluated quantities $\delta c_{\mathrm{NL}}$ and $\delta c_{\mathrm{SD}}$. In most cases, the amplitude dependence is approximated by a linear function [24], but amplitude dependencies of higher complexity have been observed as well [20,21,27,28].

Usually, this added complexity is approximated by increasing the degree of the polynomial fitting function (e.g. a quadratic function with two parameters) and study the temporal evolution of the fitting parameters, i.e., analyzing the fitting parameter evolution obtained from successive NRUS experiments. The limitation of this approach is that quadratic polynomials may not be fitting consistently well all the curves and that the resulting fitting parameters are not independent. As a result, the accuracy of the procedure is strongly compromised, and an alternative approach is needed.

Here, we will follow a novel data-based approach and introduce normalized phenomenological functions $f_{\mathrm{X}} : \mathbb{R} \to \langle 0, 1 \rangle$, which will allow us to keep the complexity of the amplitude dependence and, at the same time, to describe the evolutional property of the material with a minimal set of parameters $K_{\mathrm{X}}(t)$. In general, we can write the amplitude dependencies of each NRUS sequence collected at time $T$ (i.e., measurements performed in the time interval $T < t < T + t_{\mathrm{exp}}$ where $t_{\mathrm{exp}}$ is the duration of NRUS measurement) as

$$\delta c_{\mathrm{NL}}(\varepsilon, T) = K_{\mathrm{NL}}(T) f_{\mathrm{NL}}(\varepsilon, T), \tag{4}$$

$$\delta c_{\mathrm{SD}}(\varepsilon, T) = K_{\mathrm{SD}}(T) f_{\mathrm{SD}}(\varepsilon, T), \tag{5}$$

$$1/Q(\varepsilon, T) - 1/Q_0(T) = K_Q(T) f_Q(\varepsilon, T), \tag{6}$$

$$\Phi(\varepsilon, T) - \Phi_0(T) = K_\Phi(T) f_\Phi(\varepsilon, T). \tag{7}$$

Here $K_{\mathrm{X}}$ are scalar coefficients, while $Q_0$ and $\Phi_0$ are the quality and peak asymmetry factors at the lowest probing strain.

The definitions above are equivalent to introducing a different phenomenological function $f_{\mathrm{X}}(\varepsilon)$ for each state of the material during monitoring (each $T$). Their evolution can be approximated by a linear combination of a finite number of linearly independent components $f_{\mathrm{X},i}(\varepsilon)$, so that the $T$ dependence is contained only in the weights of such components:

$$K_{\mathrm{X}}(T) f_{\mathrm{X}}(\varepsilon, T) \approx \sum_i K_{\mathrm{X},i}(T) f_{\mathrm{X},i}(\varepsilon). \tag{8}$$

The advantage of Eq. (8) is that the time dependence (nonlinearity dependence on microstructural state) is given only by the coefficients $K_{\mathrm{X},i}$, while the amplitude dependence is described by a fixed set of functions $f_{\mathrm{X},i}$. Such approximation is possible by applying Singular Value Decomposition (SVD) to experimental data as shown later in the text and in Appendix C.

## 3. Materials and experimental setup

### 3.1. Gallium-induced embrittlement

As a general background, we briefly discuss the process of gallium-induced embrittlement as described in the literature. When an aluminum alloy is exposed to melted gallium, the liquid metal penetrates the microstructure along the grain boundaries [4]. The impact on the mechanical properties of the material is substantial. Tensile strength and elongation of the material decrease as the weakened grain boundaries prevent the full development of plastic deformation [1]. The failure mechanism changes from ductile fracture to intergranular decohesion [29].

The process of gallium penetration is complex and depends on external and internal variables of the system, e.g., external and residual stresses. The gallium wetting front is an atomic monolayer [30] but the gallium layer in between grains increases its thickness in time up to the micrometer range, following a specific grain boundary configuration and forming propagation channels [31]. The diffusion mechanism is also present, and it is accelerated by the dislocation-pipes, promoting the gallium to enter the grains. In case there is a limited reservoir of gallium, the evolution of gallium layer thickness reaches its maximum and then starts to decrease as the gallium diffuses into grains (see micrographs in [5]). Such process results in partial recovery of material mechanical properties [32].

### 3.2. Materials and damaging protocol

The material studied here is an aerospace-grade aluminum alloy 7075 in the form of a cold drawn round bar. The cylindrical sample, with dimensions of 120 mm in length and 12 mm in diameter, contained a small hole with the diameter of 1.6 mm and the depth approximately of 6 mm, drilled in the radial direction in the middle of the sample.

To induce damage, a drop of liquid gallium (approx. 1 mg) was deposited into the hole, solidifying almost immediately on contact with the cold sample surface. The sample with the applied dose of solid gallium was placed in a climate chamber (Memmert, 108L) at 20 °C so gallium remains solid. Once the temperature of the sample stabilized, the thermostat was set to 35 °C to promote the melting of gallium, which starts when it reaches 30 °C. To monitor the evolution of gallium-induced damage, NRUS probing was repeated at defined intervals for the duration of 20 h, maintaining the sample in controlled climate conditions.

### 3.3. NRUS monitoring setup and processing

The NRUS sequence described in Section 2.2 was repeated in time to track, first, the temperature transition from 20 °C to 30 °C (before beginning of gallium melting) and later the gallium damage development. The process was not sampled uniformly in time, i.e., the NRUS probing was more frequent when the parameter development was more rapid. In total, more than 200 complete NRUS sequences were collected. As a result, the experiment discussed here allows to monitor the changes in time of both linear indicators and to derive the strain dependencies of nonlinear quantities defined in Eq. (4)–(7), during the induction of gallium embrittlement. From the latter, some nonlinear indicators will be introduced and their temporal evolution analyzed as well.

The measurement of the NRUS sequence was conducted probing the first longitudinal mode of the bar. Two piezoelectric disks (height 3 mm, diameter 40 mm) were attached to the ends of the bar by a cyanoacrylate adhesive. The measurement of each NRUS sequence was performed with a TiePie HS5 oscilloscope–generator with a sampling rate of 20 MSa/s controlled by Matlab. The piezoelectric disks attached to the sample ends were used for excitation and sensing. The excitation amplitude voltage was increased up to 80 V by using a Pendulum F20AD amplifier. Linearity of both bonding and acquisition system was verified before the experiment and is evident in the measurements taken before gallium melting.

It needs to be noted here that the experimental setup had an influence on the measured linear parameters of the material. The added masses at the ends of the bar have decreased the resonance frequency of the system with respect to the expected resonance of the bare bar and led to a moderate increase in the observed damping of the system. Furthermore, the asymmetry of resonance curves in the nonlinear regime might have further influence on the definition of the $Q$ factor (and for sure on its quantification), even though this is normally not considered an issue in the literature. Therefore, in order to avoid confusion with expected material parameters of aluminum alloy, we will not use the term $Q$ factor in the following and define a normalized bandwidth $B = W/f_{\mathrm{r}}$ as a measure of damping of the system (referred to as only bandwidth further). Consistently with Eq. (6) we define evolution of bandwidth as

$$B(\varepsilon,T) - B_0(T) = K_B(T)\, f_B(\varepsilon,T)\,. \tag{9}$$

For each amplitude within the NRUS sequence, a tapered chirp excitation was used to sweep over the resonance peak. Apart from harmonic excitation, chirp signals offer the possibility of shorter measurement durations. When the sweep rate is sufficiently slow relative to the decay time, the system response approaches quasi-steady-state conditions, allowing accurate identification of resonance frequencies, while transient contributions may affect the reliable quantification of the quality factor. The bandwidth of the chirp spectra was kept constant at 2.5 kHz as was the duration of 50 ms, which was optimized to allow accurate resolution of the resonance peak at the minimal duration of excitation as discussed in Appendix A. Due to the significant change in wave velocity during the span of monitoring, the chirp center frequency was updated after each sequence to follow the changing resonance. The excitation was performed at 11 distinct amplitude steps (interleaved with baseline amplitude excitations). The whole measurement sequence was precisely timed and took 6 s to complete, i.e. each NRUS sequence had the same duration and each probing had a fixed temporal length with pauses between successive sweeps carefully controlled to compensate for delays associated with instrumentation and other factors. This accurate timing control is fundamental for a correct evaluation of $\delta c_{\mathrm{SD}}$.

For each response signal, a discrete Fourier transform with 2 Hz spacing was computed, followed by a deconvolution of the chirp spectrum, resulting in the elimination of spectral ripples and normalization of spectral amplitudes. This normalization ensured alignment between the spectral amplitude and the amplitude of the observed signal. The resonance peak was then described by several parameters (see Fig. 1): resonance frequency and amplitude, from which $c(\varepsilon)$ (and baseline velocity) is derived, peak width at half maximum from which $B(\varepsilon)$ is derived and peak asymmetry from which $\Phi(\varepsilon)$ is derived. Note that since the experiment is not calibrated, we report output amplitudes $A_{\mathrm{out}}$ in voltage (arbitrary units), which is proportional to the strain amplitude.

## 4. Results and data analysis

The results presented here refer to monitoring the gallium penetration in one sample. The repeatability of the experiment, at least for what concerns the main qualitative features discussed here, is good (see Appendix B) despite the intrinsic difficulty in obtaining a consistently equivalent gallium deposition for all tested samples.

For each NRUS sequence, a vector of velocity $c$, baseline velocity $c_{\mathrm{BL}}$, bandwidth $B$ and peak asymmetry $\Phi$ as a function of strain is collected. Using these quantities (and $c_{\mathrm{EX}}(t)$ derived as described above) we can perform the decomposition of relative velocity changes using Eq. (3) and evaluate changes in damping and peak asymmetry factors via Eq. (6) and (7). Repeating the measurement in time adds a second (temporal) dimension to the dataset, resulting in a set of matrices, where the rows correspond to amplitude steps and the columns to measurement instances.

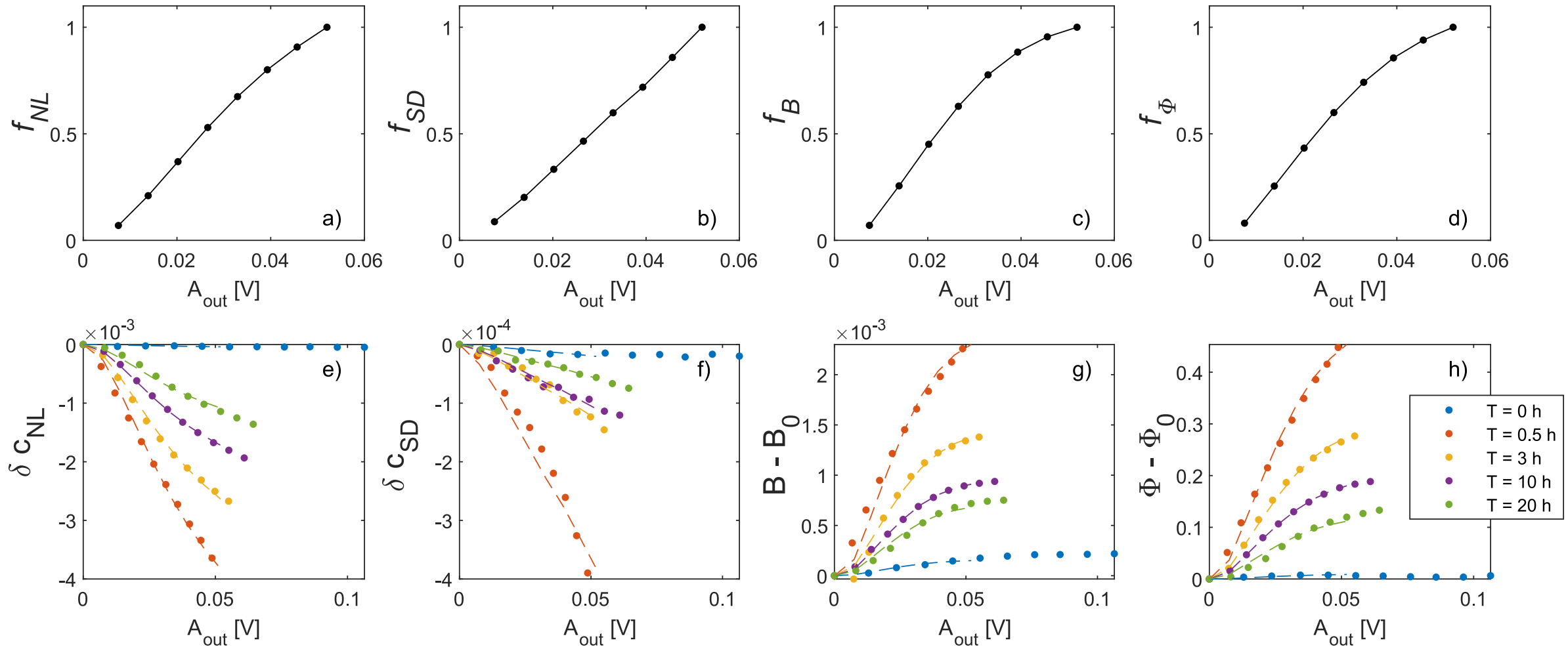

**Fig. 3.** The first row (a–d) shows the resulting phenomenological functions, i.e., the most significant singular vectors derived from the SVD approach, applied for each NRUS quantity. In the second row (e–h), examples of NRUS measurement results at different levels of gallium damage (each curve corresponds to a single NRUS sequence). The measured data are shown as dots, the fitting with the respective phenomenological function is shown as a dashed line. Note that while different strain amplitudes were achieved at varying levels of damage, the phenomenological functions all encompass the same range (see Appendix C on data processing).

### *4.1. Singular value decomposition*

The objective is to determine suitable phenomenological functions that describe the amplitude dependence of each measured quantity. To this end, the Singular Value Decomposition (SVD) was applied to the dataset. Prior to this, several pre-processing steps were performed, including resampling the data to ensure constant output amplitude intervals and normalizing each row of the data matrix (see Appendix C for details).

SVD allows us to determine the optimal and minimum number of phenomenological functions $f_{X,i}(\varepsilon)$ and the corresponding singular values $K_{X,i}(T)$ in Eq. (8), which allow describing the full-data set (strain dependences at each successive state). As discussed in Appendix C, for each of the analyzed quantities we found that it is sufficient to use a single significant singular vector $f_X(\varepsilon)$, describing the strain amplitude dependence and the corresponding vector of singular values $K_X(T)$, describing the temporal dependence. This means that we can reduce Eq. (4) to

$$\delta c_{NL}(\varepsilon, T) \approx K_{NL}(T) f_{NL}(\varepsilon). \tag{10}$$

This is of course an approximation, but it is acceptable in our context since the shape of the amplitude dependence function is not changing significantly during the course of the monitoring. We further verified that including additional singular vectors does not lead to a significant reduction in the fitting error (see Appendix C).

In the same way as $\delta c_{NL}(\varepsilon, T)$, all the other nonlinear indicators of NRUS monitoring share the same property, i.e., they have only one significant singular vector. The corresponding singular vectors bring the information on the strain amplitude dependence $f_X(\varepsilon)$ and on the temporal evolution $K_X(T)$. Note that the quantity $K_X(T)$ does not have the meaning of slope, as when data are analyzed with the usual linear fitting, but it is related to the maximum value of the corresponding property (recall that $f_X$ is a real function with values between 0 and 1).

### *4.2. NRUS indicators strain dependence*

The overview of resulting strain amplitude dependencies is shown in Fig. 3, where $f_X(\varepsilon)$ are plotted vs. output amplitude. These curves represent the nonlinearity of the sample independently from its damaged state, since the same function $f_X(\varepsilon)$ applies to data at each time. To appreciate the goodness of fit, data are shown for selected NRUS sequences, taken at different time instances during gallium damaging. The quality of the fit is consistent, with some errors for the slow dynamic contribution, which will be discussed later. The different results at different damage states are recovered by the different scaling factors applied for fitting $K_X(T)$. The strain dependence of velocity (both nonlinear and baseline), bandwidth and asymmetry (shown in the lower row of Fig. 3) indicate that gallium embrittlement initially induces a significant increase of nonlinearity, while nonlinearity diminishes at later times. The observation will be clearer when discussing the behavior of the indicators in the next Subsection.

The form of $f_X(\varepsilon)$ shows that linear fitting is not appropriate for $\delta c_{NL}$, $B$ and $\Phi$ (as anticipated) and that at the same time these functions share a common concavity and similar behavior. It is notable that the amplitude dependence of $\delta c_{NL}$ observed here is consistent with results usually obtained for consolidated granular materials [21,33], suggesting a common physical origin.

The dependence of $\delta c_{SD}$ is different since it describes the cumulative effect of conditioning of the material. Slow dynamics related effects follow an approximately linear function, which is consistent with [34]. Note that the shape of the phenomenological functions is to some extent affected by the efficiency of separation of fast and slow dynamic effects. To further discuss the properties of the NRUS procedure used here in particular for the slow dynamics contribution, a dedicated experiment was performed and the results given in Appendix D.

### *4.3. Temporal evolution of indicators*

As a result of the SVD analysis, the obtained singular values correspond to the vectors of coefficients $K_X(T)$ (for nonlinearity, slow dynamics, $B$ and $\Phi$), which allow to map the evolution of nonlinearity in time, i.e., during the evolution of gallium–aluminum interaction. In addition, the linear velocity $c_0$ and linear $B_0$ are also measured in each NRUS sequence, thus we can follow their time dependence.

The evolution of all indicators in time is shown in Fig. 4 and shares common characteristics. Let us first recall that the initial part of the evolution (0 – 0.4 h) describes the phase during which the gallium deposited on the sample remains solid, while the temperature changes from 20 °C to 30 °C and it is the most sensitive to the procedure of gallium dropping in the specimen hole. Even during this preliminary phase, we suspect to already have a small partial gallium penetration (see Appendix E).

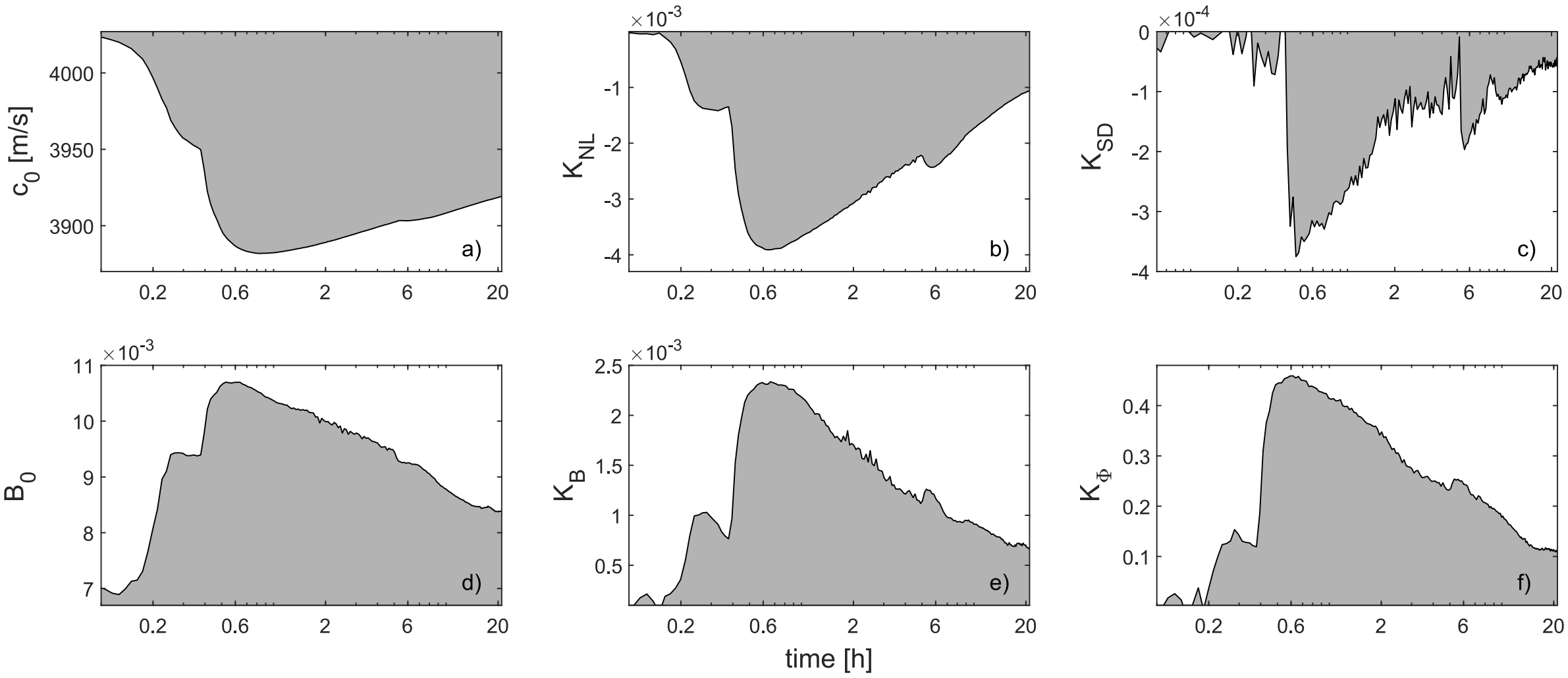

**Fig. 4.** Evolution of linear and nonlinear elastic parameters during gallium damaging. Temporal evolution of (a) linear elastic wave velocity, (b) coefficient of nonlinear elasticity, (c) coefficient of slow dynamics, (d) linear damping, (e) coefficient of nonlinear damping, (f) coefficient of asymmetry. Note that plots are shown in logarithmic time scale.

After approx. 0.4 h, we observe a sudden and discontinuous variation in the evolution, corresponding to the time instance at which the temperature of 30 °C is reached in the sample and gallium rapidly melts. From here, nonlinear indicators evolve very rapidly (decrease) until a minimum is reached, after 0.6 h. This indicates a complete melting of the gallium dose and the end of the phase of fast penetration into the microstructure of the material.

At later times, nonlinear indicators show a slow recovery process. Note that nonlinearity does not recover its initial value in the time window of observation. This phase is consistent with the second phase of gallium penetration: leaving the grain boundaries and diffusing into the grain bulk. Some gallium is also expected to remain trapped on the sample surface and that permanent damage will remain impacting elastic properties of the samples. At about 6 h, the presence of an anomalous peak can be observed, consistently for all indicators. This effect was also observed in one of the repeated experiments, even though not as strongly (see Fig. B.1 in Appendix B). Therefore, it likely represents some change in the gallium diffusion process, but at the current stage we are unable to ascribe it to any known mechanism as such long-term observations were not performed before.

Finally, let us remark that the evaluation by NRUS is not local, i.e., the measured properties are a conjoined result of the distribution of probing strains along the length of the sample and of gallium traveling from the center of the sample towards its ends.

We can conclude that nonlinear indicators provide a sensitive and reliable tool to monitor what is physically happening within the structure. In particular, they can be used as an indirect tool to determine relevant times within the process, e.g. the time at which gallium starts infiltrating in the grain boundaries and the transition from grain boundaries infiltration to grain bulk diffusion, which was also observed in the literature (see micrographs in Fig. 3 [31] showing a change in the evolution of gallium thickness at grain boundaries after about 8 min or in Fig. 4 of [5] showing migration of gallium away from grain boundaries at very late times). Other micrography observations partly confirm this evolution of LME in aluminum [35].

## 5. Discussion

### 5.1. Linear indicators

Let us start discussing the linear indicators (velocity and $B_0$ factor), shown in Figs. 4a and 4d. The precision of the method to resolve relative changes in wave velocity is of the order of $10^{-5}$, (useful for what concerns the nonlinear indicators analysis) and exact. Thus, the velocities variations, both as a function of time and strain are accurately predicted. While the curves observed in Figs. 4a and 4d are reliable in shape, the absolute values of velocity calculated from the lowest amplitude resonance curve present a significant offset with respect to the expected values of 6300 m/s, reported in the literature for aluminum alloys. This is due to the fact that the absolute values of wave velocity are derived from the resonant frequency of the system assuming free–free boundary conditions, but neglecting the added mass of the piezoelectric disks attached to the ends of the bar [36]. A Comsol simulation of the entire system (including transducers) confirmed that the observed discrepancy in linear velocity was due to such an effect, but also confirmed the proportionality between measured and real values, thus ensuring that results discussed here are meaningful. The evaluation of the $B$ values is also dependent on the boundary conditions, thus the quantity $B^{-1}$ is not equivalent in absolute value to the material $Q$ factor, as already discussed. Nevertheless, it still conveys information about damping evolution. Indeed, in the literature it is common to find the 'normalized' bandwidth assimilated to $Q^{-1}$.

The temporal evolution of velocity and damping (through the bandwidth) still conveys reliable and meaningful observations about damage evolution. From the data reported in Fig. 4, we notice that in the phase before reaching the maximum variation of the linear parameters, the behavior is very similar to that observed for nonlinear indicators (compare Fig. 4a and d with Fig. 4b, c, e and f). In contrast, in the recovery part of the curve, linear indicators show a significantly slower recovery speed compared to the analogous recovery of nonlinear indicators, together with more incomplete recovery. That results in a lower sensitivity in the detection of the time corresponding to the beginning of recovery.

### 5.2. Nonlinear indicators

For what concerns the nonlinear indicators, the coefficients of nonlinear elasticity $K_{\mathrm{NL}}$, nonlinear damping $K_B$ and fast dynamics $K_\Phi$, all show a similar temporal evolution. The linear correlation between $K_{\mathrm{NL}}$, $K_B$, $K_\Phi$ during the whole process of gallium damage (see Fig. 5) means that variations of modulus, damping and asymmetry stem all from a common physical origin, similarly to what happens in the conditioning and relaxation phases [37]. This is consistent with the concept of nonequilibrium strain presented in [38]. The linear correlation between the change of peak asymmetry and $B$ factor also suggests that, likely, the amplitude dependent change in damping is mostly due to the increasing peak asymmetry, i.e., peak tilting (see Fig. 5a), which makes the interpretation of width at half maximum more troubling. It has been shown that the tilting of the peak observed at high excitation amplitudes is a result of a measurement protocol rather than a specific material model [39]. Therefore, an alternative definition of damping might be necessary, as also discussed in [40].

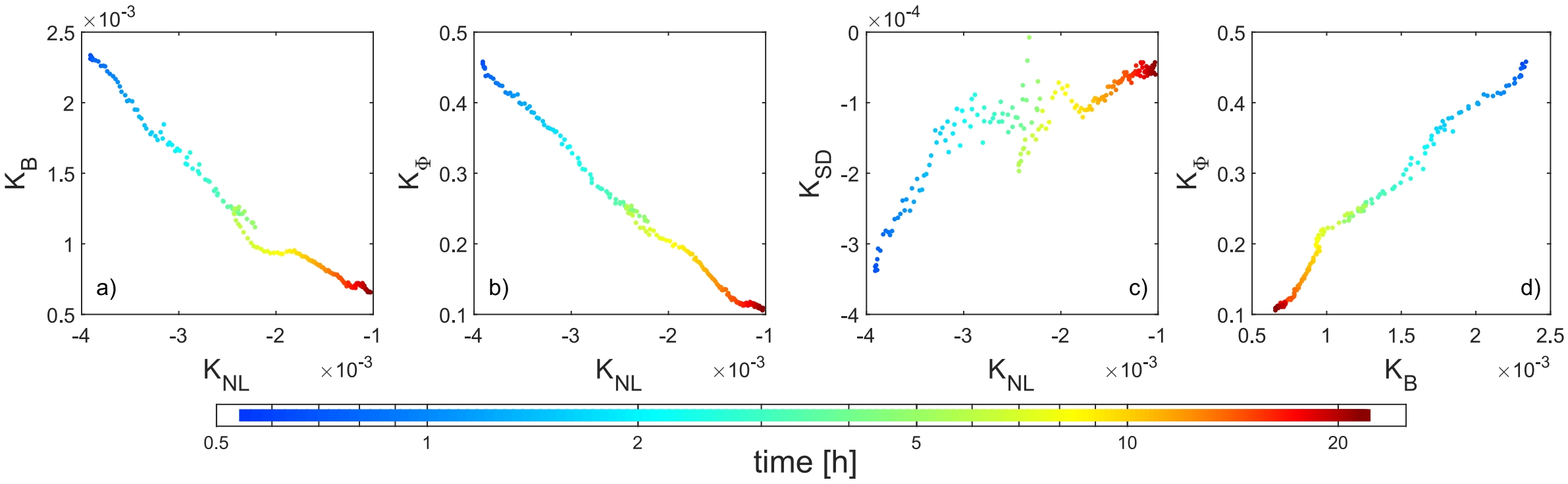

**Fig. 5.** Correlation of nonlinear indicators. (a–c) Bandwidth (damping), asymmetry, and slow dynamic coefficients as a function of nonlinear velocity coefficients. (d) Asymmetry coefficients as a function of damping coefficients. The color indicates the time of measurement. (For interpretation of the references to color in this figure legend, the reader is referred to the web version of this article.)

The temporal evolution of the slow dynamic coefficient $K_{SD}$ is different and it does not correlate with the velocity indicator $K_{NL}$ (see Fig. 5c). As it can be seen in Fig. 4c, the $K_{SD}$ coefficients are an order of magnitude lower than $K_{NL}$. Thus, noise present in the data becomes more relevant (lower signal-to-noise ratio). Also, some issues might arise due to the determination of $c_{EX}$, which again might be affected by some noise (i.e., not based purely on measured data) and involve approximations due to data interpolation, which might be of the order of magnitude comparable with the effects on baseline velocity variations. Both effects make the estimate of $K_{SD}$ less robust than for the other indicators.

Nevertheless, the lack of linear correlation between $K_{SD}$ and $K_{NL}$, might also have a reasonable physical interpretation. In fact, slow dynamics is a long-time process and the repetitive NRUS probing does never allow full relaxation. The resulting amplitude dependence $f_{SD}$ and temporal evolution of $K_{SD}$ could thus be consequence of a cumulative conditioning in the presence of incomplete relaxation. We could expect that, when cumulative effects are accounted for (either in specific experiments or more accurate postprocessing), $K_{SD}$ might correlate linearly with the other $K_X$ values as well. But this is not shown by our data.

## 6. Conclusions

We have applied Nonlinear Resonant Ultrasound Spectroscopy (NRUS) to monitor the diffusion of liquid gallium into an aluminum specimen, inducing Liquid Metal Embrittlement (LME). Indicators introduced to measure elastic nonlinearity (linked to both fast and slow dynamics) evolve in time and follow the typical phases of gallium diffusion along grain boundaries (in a first phase) and within the bulk of the grains (in a second phase). A transition between the two behaviors is determined, corresponding to a variation in the trend of the nonlinear parameters and monitoring nonlinear parameters looks more sensitive in determining such a transition than monitoring linear velocity or linear damping.

The use of NRUS to monitor LME has several advantages, because the procedure is easy to implement and very sensitive to microstructural changes in the material. Furthermore, it can be easily implemented for long-time monitoring. The limitations are mostly due to the fact that it provides averaged information over the entire specimen volume, thus it is not particularly suitable for localization, even though a few modifications of the procedure were introduced to allow it [41]. In this work, we also aimed to prove benefits of using nonlinear probing. In the case reported here, the advantages with respect to probing linear resonance evolution are minimal, but still influent in the sense they allow better sensitivity in the determination of the inversion time between softening and hardening (the minimum in Fig. 4 is sharper for nonlinear indicators). We also remark that the feasibility of a nonlinear analysis could be relevant because nonlinear effects are known to be more sensitive than linear ones to small changes in the microstructure (e.g. in the case of embrittlement due to very tiny amount of gallium). Also, the very long term evolution of indicators ($t \to \infty$, not studied here) might be different, with implications for the characterization of the final state of the material (likely full recovery of nonlinear indicators and partial recovery of linear ones).

The analysis is made complex by the embrittlement of the material, so the nonlinear behavior is changing over time not only quantitatively but also in its physical interpretation. As a consequence, data analysis required to introduce a novel approach, based on the Singular Value Decomposition (SVD) method, which revealed a particularly efficient way of identifying common (and therefore quantifiable and comparable) features of the sample in different damage states.

The SVD approach is general and could be applied to other experimental conditions, whereas it is needed to monitor variations of specific material properties independently of variations due to other parameters, which are varying as well. Further work will also be performed to assess the link between the evolution of embrittlement (transient phase of damage propagation) and the resulting final damage (equilibrium phase). Both could indeed be measured with NRUS, allowing one to identify parameters governing the transient damage process that control the final damage state.

The correlations between the evolution of the relative velocity change and other nonlinear indicators are consistent with recent observations [21,42] and modeling efforts [43,44] highlighting the importance of fast and slow dynamics in the nonlinear response of certain materials. The gallium damaging further reinforces the hypothesis that these phenomena stem from the interaction of stiff grains with softer, viscoelastic intergrain matrix. This is also confirmed by the gradual diminishing of nonlinearity, as gallium leaves the intergrain volume and diffuses into grain bulk.

## CRediT authorship contribution statement

**Jan Kober:** Writing – review & editing, Writing – original draft, Visualization, Validation, Supervision, Methodology, Investigation, Data curation, Conceptualization. **Radovan Zeman:** Writing – review & editing, Visualization, Methodology, Investigation. **Josef Krofta:** Writing – review & editing, Visualization, Validation, Investigation, Formal analysis, Data curation. **Antonio S. Gliozzi:** Writing – review & editing, Methodology, Formal analysis, Conceptualization. **Marco Scalerandi:** Writing – review & editing, Validation, Supervision, Methodology, Investigation, Formal analysis, Conceptualization.

## Acknowledgments

J. K. and R. Z. acknowledge the financial support provided by the Ministry of Education, Youth, and Sports of the Czech Republic via the project No. CZ.02.01.01/00/23_020/0008501 (METEX), co-funded by the European Union. J. Kr. is supported by the Academy of Sciences of the Czech Republic within the Programme AV 21 – Breakthrough Future Technologies. J. Ko., J. Kr. and R. Z. are funded by the institutional support through the grant RVO: 61388998.

## Declaration of competing interest

The authors declare that they have no known competing financial interests or personal relationships that could have appeared to influence the work reported in this paper.

## Appendix A. Chirp duration

As discussed in the main text, the resonance curve obtained using chirp excitation is equivalent to that obtained by frequency swept continuous waves only when the chirp duration $t_{chirp}$ is sufficiently long to guarantee the sample is consistently close to standing wave conditions. In that case, the resonance frequency is correctly estimated, while some quantitative errors in the estimation of the $Q$ factor from the bandwidth $B$ might still be present. Therefore, a careful calibration was performed before starting the experiment.

The sample was tested before any gallium treatment, and several experiments were performed varying $t_{chirp}$ and calculating the resonance frequency and $B$. The results are plotted vs. chirp duration in Fig. A.1 (blue circles) and 10 repetitions at each value of $t_{chirp}$ were performed. We choose ($t_{chirp}$ = 50 ms), because it is the shortest duration which made possible to measure reliably the $B$ parameter. Recall that after gallium embrittlement, damping increases thus guaranteeing the choice made is adequate also for samples monitored during embrittlement.

Second, we repeated the same test using a laser vibrometer as a detector instead of the PZT, to allow better calibration of the detected signal. Results (red squares in Fig. A.1) are in good agreement with those obtained from PZT.

## Appendix B. Repeatability

Repeatability provides robustness of the approach. However, in the tested case the bottleneck in demonstrating it lies in the procedure of gallium deposition, which is hard to repeat with the very same conditions due to environmental parameters and the procedure of transfer of samples from the external environment to the climate chamber. As a result, more variable results are expected for some of the tested samples and/or part of the initial evolution might be lost because of imperfect gallium solidification within the hole immediately after deposition. Given such consideration, we expect only qualitative repeatability of the behavior, with the main features observed for all tested samples, while some of the details might be lost.

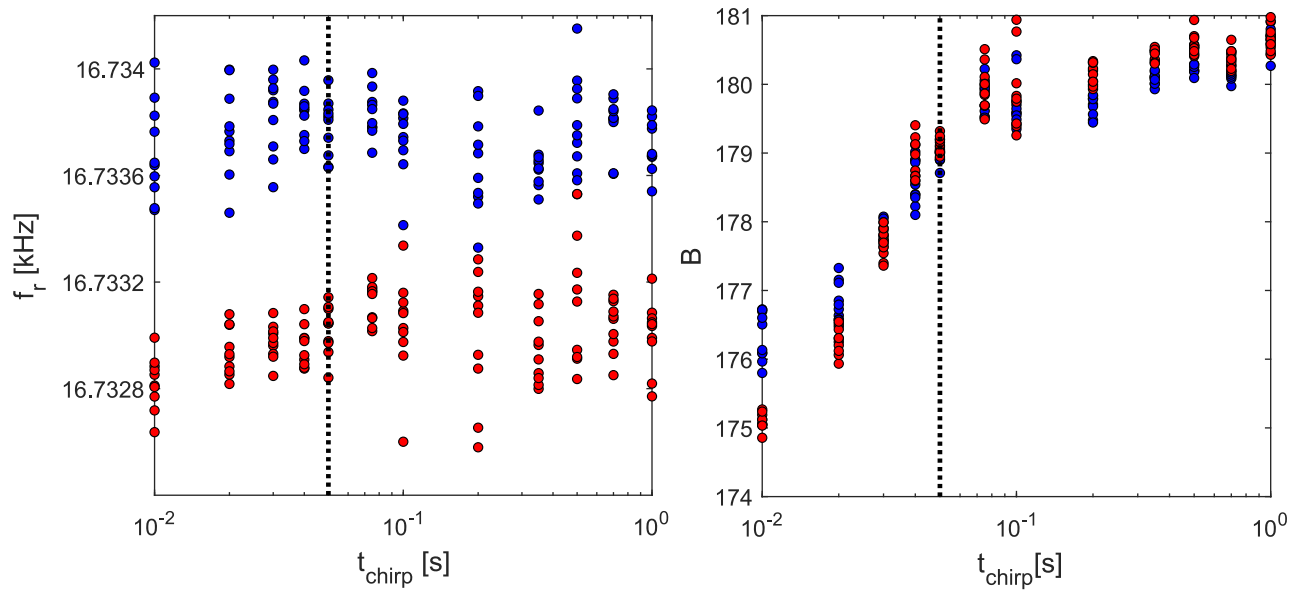

**Fig. A.1.** Resonance frequency and bandwidth vs. chirp duration using as receiver a piezoelectric disk (blue) or a Laser Doppler Vibrometer (red). (For interpretation of the references to color in this figure legend, the reader is referred to the web version of this article.)

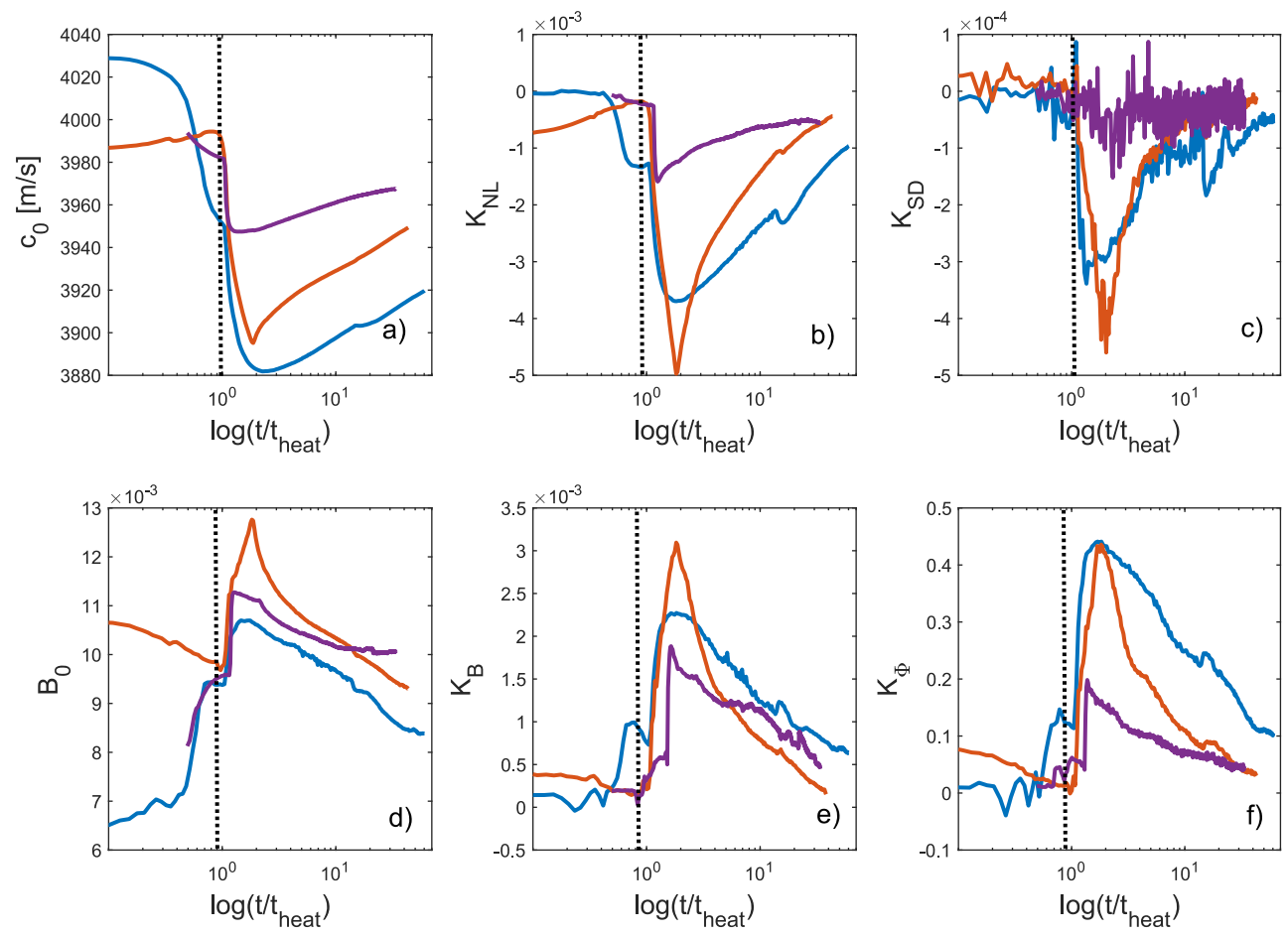

**Fig. B.1.** Evolution of linear and nonlinear elastic parameters during gallium damaging for three tested samples (different colors). Temporal evolution of (a) linear elastic wave velocity, (b) coefficient of nonlinear elasticity, (c) coefficient of slow dynamics, (d) linear damping, (e) coefficient of nonlinear damping, (f) coefficient of asymmetry. Note that plots are shown in logarithmic $x$ axes scale, time $t$ normalized to end of heating time $t_{heat}$ (estimated). The vertical black dashed lines separate the heating phase from the LME phase. (For interpretation of the references to color in this figure legend, the reader is referred to the web version of this article.)

Within the consideration above, to study repeatability other samples were prepared and tested following ideally the very same approach (including timing of measurements). Among the main issues which we have identified during preparation, the following are relevant for the interpretation of results in the heating phase (before gallium melting within the climate chamber, i.e., in the first 0.4–0.6 h):

- the initial temperature of the climate chamber was kept at 20 degrees, while the environmental temperature was not controlled, thus it is expected that the temperature of the sample when entering the climate chamber was different for different cases. If it was larger than 20 degrees (and smaller than 30 degrees, of course), it may result in losing at least partly the evolution during heating (first 0.4–0.6 h);
- the amount of gallium deposited was only qualitatively controlled. A slightly larger weight of the gallium droplet and/or a non-perfect insertion in the bottom of the hole might lead to very tiny amounts of gallium not solidifying instantly when in contact with Al. This might result in some evolution of the nonlinear parameters already during the heating phase due to an initial/weak penetration into grain boundaries (as observed for sample 1 analyzed in the paper and reported in Appendix E).
- the difference in the amount of gallium deposited together with the local different microstructure of the specific sample might favor higher/lower amount of gallium penetration, resulting in absolute values of the parameters variation which might be very different (e.g. sample 3);

We also observe that the testing strain amplitudes for different samples span slightly different ranges, even though the same drive amplitudes were used. That is caused by slightly different boundary conditions and quality of the bonding.

The overall results obtained for all samples support the previous findings: the development of linear and nonlinear parameters shares the same behavior, presenting a slow recovery of viscoelastic properties (Fig. B.1), while keeping the linear correlation observed in the first experiment. We also remark that the slow dynamic indicator $K_{\mathrm{SD}}$ seems the one with both highest repeatability and better sensitivity to both gallium penetration onset and transition to the phase in which it is dispersed into grains.

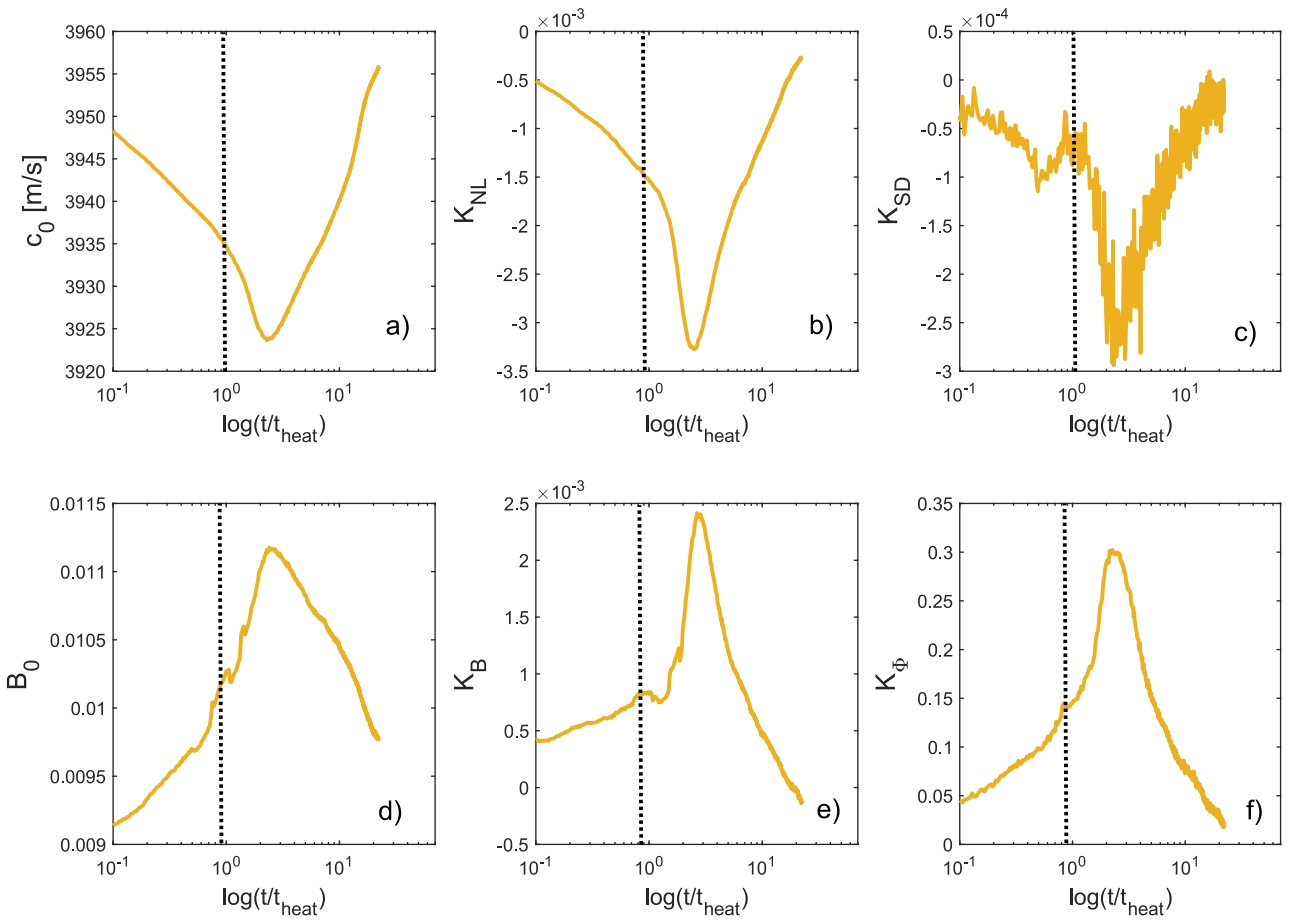

**Fig. B.2.** Evolution of linear and nonlinear elastic parameters during gallium damaging for sample 2 applying a second gallium dose few days after the first one. Temporal evolution of (a) linear elastic wave velocity, (b) coefficient of nonlinear elasticity, (c) coefficient of slow dynamics, (d) linear damping, (e) coefficient of nonlinear damping, (f) coefficient of asymmetry. Note that plots are shown in logarithmic $x$ axes scale, time $t$ normalized to end of heating time $t_{heat}$ (estimated). The vertical black dashed lines separate the heating phase from the LME phase.

It is remarkable that the same features in the evolution are also shown when a dose of gallium is applied to a sample which was already subject to embrittlement. In Fig. B.2 the evolution of the indicator for sample 2 when a second dose of gallium was applied (after a few days from the first dose applications). Recall that results for the first application are shown in Fig. B.1 in red. The presence of a sharp change in the indicators at the end of the heating phase (although not as sharp as for intact samples in Fig. B.1) and their recovery after the maximum/minimum was reached are clear. The phase of gallium infiltration into grain boundaries (from end of heating to time of minimum/maximum variations) is clearly longer with respect to previous cases. A difference in infiltration speed was expected, because of the changes in the grain boundaries structure already caused by the first gallium dose.

Finally, we remark that the phenomenological functions for all samples share the same behavior and that the linear correlations among parameters exhibited by sample 1 (discussed in the text) is also a typical property of the other tested samples.

## Appendix C. Data preprocessing and singular value decomposition results

The results of NRUS monitoring come in the form of matrices, where the rows correspond to amplitude steps and the columns to measurement instances. Here we will discuss in more detail, on the example of $\delta c_{\mathrm{NL}}$, the properties of the data matrix and all the data processing steps required to evaluate the phenomenological function $f_{\mathrm{NL}}$ and the coefficients of temporal evolution $K_{\mathrm{NL}}$.

As mentioned in the main text, the objective is to identify a minimum number of phenomenological functions in Eq. (8) that optimally satisfy Eq. (4) for all datasets (at successive time instances). Some preliminary steps have to be implemented to make the data suitable for the analysis. The primary concern is amplitude sampling. NRUS data are sampled at constant input amplitude steps, but the output amplitudes (corresponding to excitation strains) reach different values at each measurement instance (Fig. C.1a), due to the temporal dependence of the viscoelastic properties of the sample. To proceed with the analysis, for each measurement instance, the $\delta c_{\mathrm{NL}}$ curve was resampled to constant output amplitude steps. The values of the output amplitudes vector were selected based on the histogram of the output amplitudes of the whole monitoring (Fig. C.1b). The Akima's bivariate interpolation was used for resampling [45], which also allowed a moderate extrapolation of some curves in order to extend the range of output amplitudes considered. The resampling worked out very efficiently, as can be seen in Fig. C.1c.

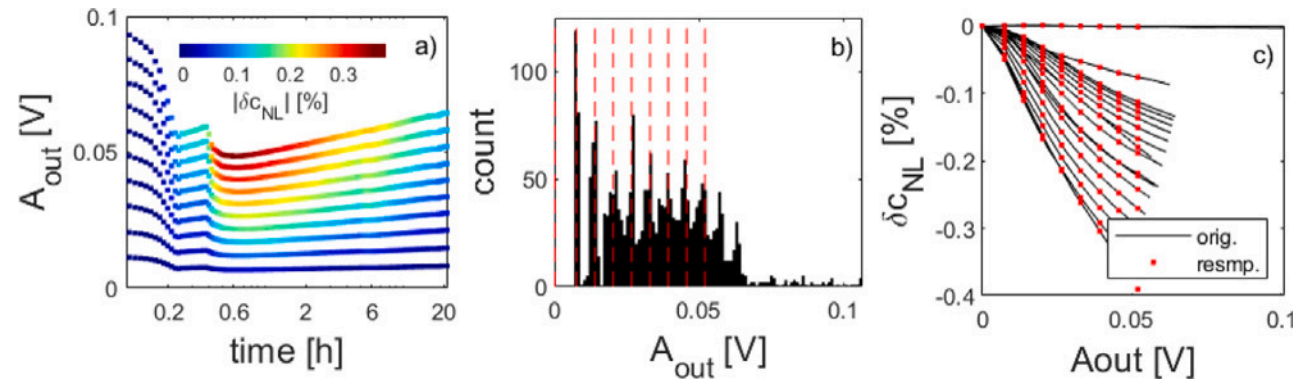

**Fig. C.1.** Resampling of the data matrix. (a) Data of relative velocity change shown in time of measurement and output amplitude coordinates. (b) Histogram of output amplitude values from the whole monitoring. Red dashed lines indicate the $A_{\mathrm{out}}$ values used for resampling. (c) Resampled values superimposed on original data for selected measurements. (For interpretation of the references to color in this figure legend, the reader is referred to the web version of this article.)

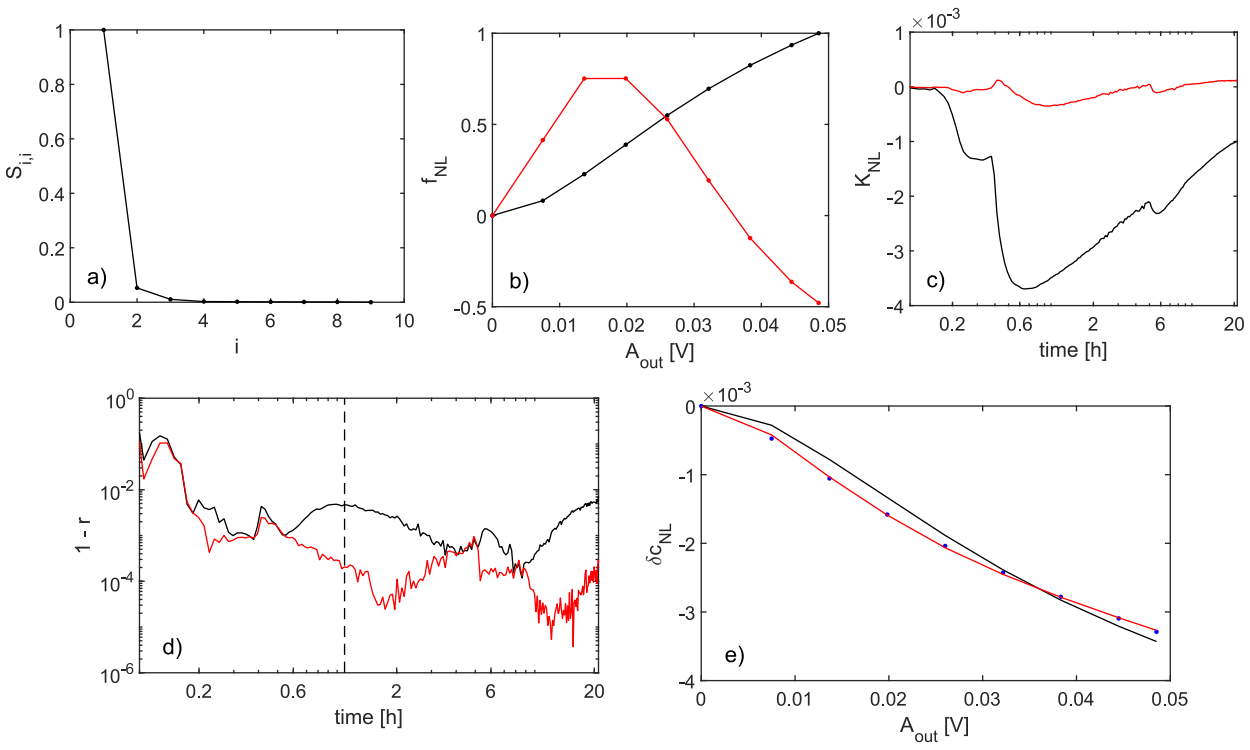

**Fig. C.2.** Components of SVD. (a) Singular values $S_{ii}$; (b) Two most relevant singular vectors; (c) Singular values as a function of time for the first two singular vectors; (d) Goodness-to-fit obtained using one (black) or two (red) singular vectors; (e) Fit of the experimental data using one (black) or two (red) singular vectors. (For interpretation of the references to color in this figure legend, the reader is referred to the web version of this article.)

Secondly, to obtain unbiased decomposition, we normalized each row of the data matrix, i.e., each NRUS amplitude sequence.

In order to find the relevant eigenfunctions $f_{\mathrm{NL},i}$, we use the Singular Value Decomposition (SVD) method, which is a technique for solving the singular value problem for rectangular matrices and allows assessment of the dimensionality of the data. Let us denote the data matrix $\delta c_{\mathrm{NL}}(\varepsilon, t)$ as $\boldsymbol{M}$. The SVD method finds the matrices $\boldsymbol{U}$, $\boldsymbol{S}$ and $\boldsymbol{V}$ such, that

$$\boldsymbol{M} = \boldsymbol{U}\boldsymbol{S}\boldsymbol{V}^{\mathrm{T}}. \tag{C.1}$$

Here $\boldsymbol{M}$ is a $m \times n$ matrix, where $n$ counts amplitude steps and $m$ measurement instances, while $\boldsymbol{S}$ is a diagonal matrix $m \times n$ with singular values arranged in descending order. The respective singular vectors are given by the matrix $\boldsymbol{V}$ $(n \times n)$ and $\boldsymbol{U}$ is a matrix $(m \times m)$ defining the coefficients of the linear combinations of singular vectors, which map the new base on the original data.

The results of the decomposition allow us to determine the singular vectors (indexed by $i$), sorted by relevance given by the value $S_{ii}$. Each NRUS curve is then interpreted as a linear combination of those singular vectors. Introducing the notation $V_{ij} = V_i(\epsilon)$, we can write:

$$\delta c_{\mathrm{NL}}(\epsilon, T) = \sum_{i=1}^{n} K_{\mathrm{NL},i}(T)\, V_i(\epsilon)\,, \tag{C.2}$$

where the coefficients $K_{\mathrm{NL},i}$ are the elements of the vector $\boldsymbol{US}$.

The analysis of the experimental data gives the relevance of successive singular vectors, as shown in Fig. C.2a. The results show that there is a dominant singular value corresponding to $V_1$, while other singular vectors ($V_i$ with $i \neq 1$) have a minimal contribution. This confirms the hypothesis of Eq. (10) in the main text: only one singular vector is relevant/sufficient to describe the observations. We might observe that a second singular vector ($i = 2$) might also have some relevance. $V_1$ and $V_2$ are shown in Fig. C.2b, even though only the main one was used in the main text.

To further assess the approximation of using only one eigenvector to fit the data and to better define the relevance of the other singular vectors, we tested the difference of goodness of fit when one or two singular vectors are used to fit NRUS data. The results of the calculations are shown in Figs. C.2d and C.2e. The goodness of fit is evaluated as a normalized correlation coefficient $r$. The fitting results yield typically $r = 0.995$ and better. Obviously, decomposition into two singular vectors provides a better fitting, but the difference is not substantial. The temporal evolution of the eigenvalues corresponding to the dominant eigenvectors is shown in Fig. C.2c.

In the main text, we have considered the approximation sufficient considering only one function describing the strain dependence. However, the analysis shown here seems to indicate that we could be more precise if two linearly independent components are used. However, in our opinion, the slight relevance of the second eigenvector is due only to the incomplete elimination of slow dynamic components. Indeed, when measuring the baseline velocity $c_{\mathrm{BL}}$ some time has passed, so we could expect that part of the conditioning effects have been relaxed. Thus, elimination of slow dynamics is not complete. This assumption seems to be confirmed when considering the temporal evolution of the coefficient which multiplies the second singular vector (red in Fig. C.2c), which resembles the evolution of the slow dynamic coefficient (see Fig. 4c in the main text).

## Appendix D. Baseline correction efficiency

The evaluation of relative velocity changes due to nonlinearity $\delta c_{\mathrm{NL}}$ is based on using low-amplitude velocity measurements as "immediate" baselines (see Eq. (4)). The objective is to remove all environmental and cumulative contributions to the velocity change. To assess the efficiency of such approach, a dedicated experiment was conducted after the gallium damage tracking was finished (i.e., at 22 h after gallium deposition). The NRUS sequence was modified to include upward and downward amplitude ramps (Fig. D.1a). Timing and excitation levels were kept the same as in the testing performed in our analysis.

The results are shown in Fig. D.1b. First, the raw velocity data are computed without using the baseline measurements (red). As expected, upward and downward branches do not match, with the downward branch showing a larger relative change. This is an effect due to progressive conditioning (slow dynamic effect). In the same experiment, the raw baseline velocities are also measured (red curve in Fig. D.1c). Plotting the difference between high-amplitude and baseline velocities (at each strain) lead to a decrease in the total nonlinearity and contributes to closing the gap between upward and downward branches (black curve in Fig. D.1b). However, the cumulative conditioning effect is not completely removed.

Thus, removing the baseline velocity variation from the total velocity variation does not remove completely slow dynamic effects, since a significant part of relaxation happens in a fraction of a second.

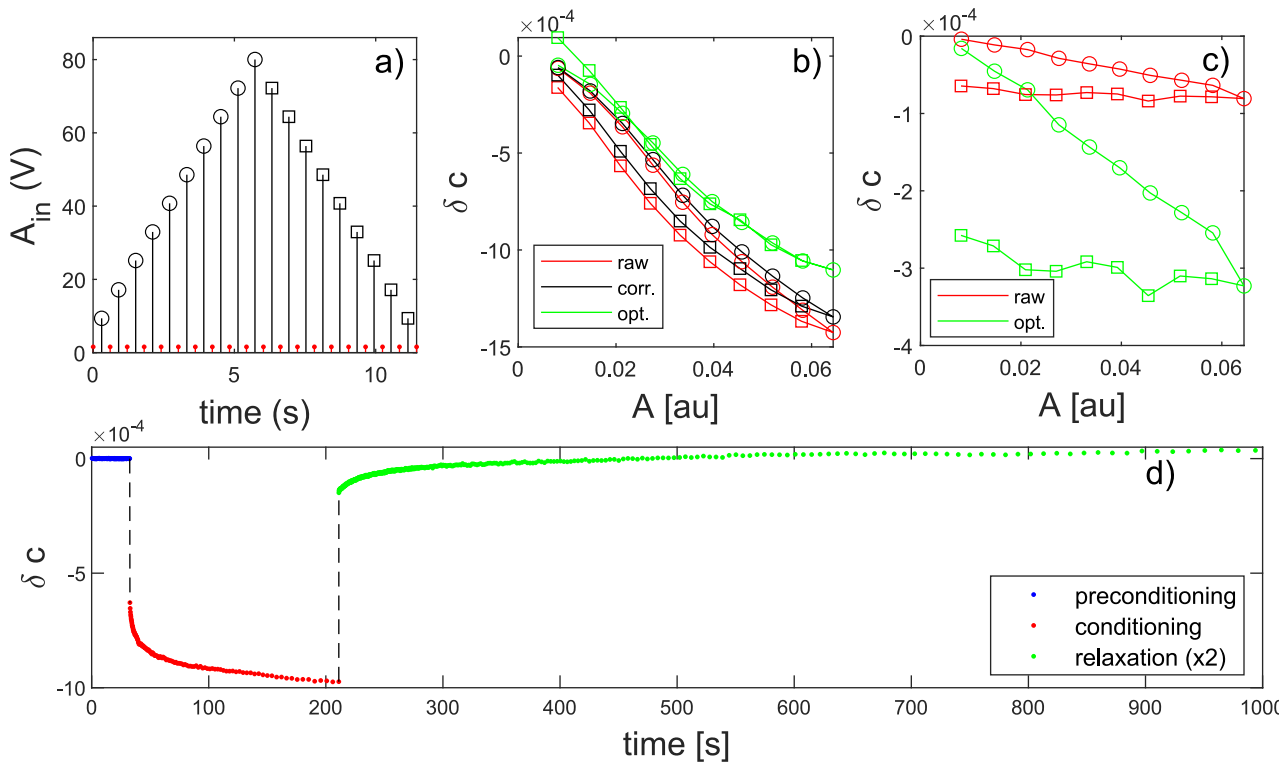

**Fig. D.1.** (a) Modified NRUS amplitude sequence including a downward ramp. The sequence timing agrees with previous probing. (b) Relative velocity change vs. strain: computed without baseline reference correction (red), with baseline reference correction (black) and with amplified baseline reference correction (green). (c) Baseline reference vs. strain: measured (red) and amplified (green) baseline velocity variation. (d) Slow dynamic response of gallium damaged sample. Measurements are taken on the first sample after 40 h of gallium damaging. Note that the relaxation curve is amplified by a factor of 2 to allow appreciating the temporal evolution. (For interpretation of the references to color in this figure legend, the reader is referred to the web version of this article.)

To prove this result, we performed a typical conditioning/relaxation experiment [46]. First the sample is probed, i.e., velocity is measured at successive times at constant large amplitude of excitation (conditioning, red curve in Fig. D.1d). After the conditioning time, velocity is measured at successive times with a low amplitude chirp excitation (relaxation, green curve in Fig. D.1d).

Since in our experiments the chirp repetition rate is 0.3 s, the baseline measurement is not able to capture the initial point of the relaxation curve. In other words, the baseline velocity is not measured at $t = 0$, i.e., at the initial instant when relaxation starts, and most of the jump from the conditioned state (end of the red curve in Fig. D.1d) has already taken place. Yet, since the sequence is timed precisely, we can attempt to retrieve and quantify the velocity variation occurred in the time needed to perform the baseline measurement.

The procedure is based on measured data. Considering the relaxation occurring in the first 0.3 s, we have estimated that about 75% of the relaxation process was achieved, before measuring baseline velocity variations. Thus, a scaling of the baseline data by a factor 4 is expected to be required, leading to a rescaled baseline curve (green curve in Fig. D.1d). Subtracting the rescaled baseline from measured velocity variations, loops are completely closed (green curves in Fig. D.1b).

In conclusion, the total effect of cumulative conditioning on NRUS data is four times larger than the baseline measurements, suggesting that the complete removal of cumulative conditioning should decrease the total nonlinear response by 20%.

## Appendix E. Evolution of linear and nonlinear parameters during heating

The initial part of the temporal evolution of linear and nonlinear parameters (Figs. 4 and B.1) monitors the evolution of viscoelastic linear and nonlinear parameters during heating in the climate chamber (from 20 °C to 35 °C), which corresponds to heating of the sample (with some delay). As mentioned before, results observed during such phase are strongly biased by the procedure of gallium deposition and the environmental conditions at which that occurs.

As an example of such influence, we focus here on discussing the unexpected sensitivity of the nonlinear coefficients to temperature variations (see the black curves in Fig. E.1 which corresponds to a zoom of the first instances of Fig. 4). In this phase, the sample was considered intact, and the presence of solid gallium deposited in the central hole should not significantly perturb the system state. Indeed, an experiment was performed on a sample without deposited gallium (red curves in Fig. E.1) with negligible effects of temperature variations on nonlinear parameters, while evident variations of the linear parameters are observed: wave velocity and bandwidth decrease with increasing temperature, as expected. The strong dependence of nonlinear parameters observed when gallium is introduced, cannot thus be due only to heating (compare black with red curves). Thus, we suppose that a small portion of gallium must have rapidly penetrated the microstructure during the deposition of the gallium on the sample, which has occurred during the short time taken by the gallium droplet to solidify on the surface. This damaging phase is of course not captured by our analysis.

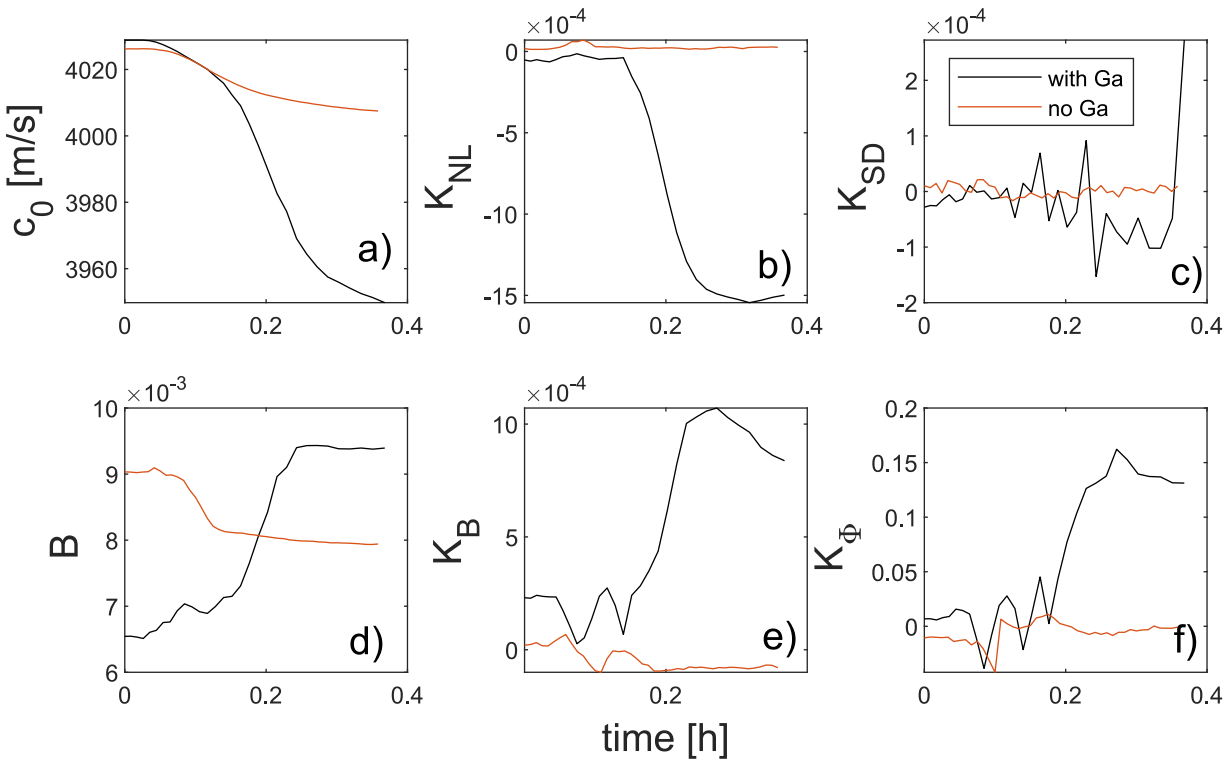

**Fig. E.1.** Linear and nonlinear elastic parameters during the heating phase, i.e., as a consequence of a temperature change, without (red) and with (black) deposited gallium. Temporal evolution of (a) linear wave velocity, (b) coefficient of nonlinear elasticity, (c) coefficient of slow dynamics, (d) linear damping, (e) coefficient of nonlinear damping, (f) distortion coefficient. (For interpretation of the references to color in this figure legend, the reader is referred to the web version of this article.)

## Data availability

Data will be made available on request.